\documentclass[brazilian,twocolumn,aps,prb,superscriptaddress,amsmath,amssymb,showpacs,floatfix,preprintnumbers]{revtex4-2}
\usepackage[latin9,utf8]{inputenc}
\usepackage{float}
\usepackage{textcomp}
\usepackage{amsmath,amssymb,amscd,hyperref}
\usepackage{graphicx}
\makeatletter
\usepackage{url}
\usepackage{grffile}
\usepackage{bbold}
\usepackage{comment}
\usepackage{dirtytalk}
\usepackage{chngcntr}
\usepackage{tabularx}
\usepackage{multirow}
\usepackage{soul}
\usepackage{cleveref}
\usepackage{braket}
\usepackage{makecell}
\DeclareUnicodeCharacter{2212}{-}
\DeclareUnicodeCharacter{0301}{\hspace{-1ex}\'{ }}

\usepackage{dcolumn}
\usepackage{color}
\DeclareGraphicsExtensions{.png .jpg .pdf}
\usepackage{hyperref}
\hypersetup{
     colorlinks = true,
     linkcolor = blue,
     anchorcolor = blue,
     citecolor = blue,
     filecolor = blue,
     urlcolor = blue
     }
\usepackage{braket}
\usepackage{physics}
\usepackage{array}
\usepackage{booktabs}
\usepackage{soul}

\makeatother
\newcommand{\ZHG}{\mathrm{ZHG}}
\newcommand{\ZHGo}{\mathrm{ZHGo}}
\newcommand{\ZHGda}{\mathrm{ZHGda}}
\newcommand{\ZHGdb}{\mathrm{ZHGdb}}
\newcommand{\SHG}{\mathrm{SHG}}
\newcommand{\ABS}{\mathrm{ABS}}
\newcommand{\TR}{\mathrm{TR}}

\begin{document}

\title{Quantum Interband-Coherence-Driven Inverse Faraday Effect in Intrinsic Graphene}

\author{X. D. Wang}
\affiliation{School of Materials and Physics, China University of Mining and Technology, Xuzhou 221116, P. R. China}
\affiliation{Center for Quantum Matter, School of Physics, Zhejiang University, Hangzhou 310027, P. R. China}

\author{H. M. Dong}
\email{hmdong@cumt.edu.cn}
\affiliation{School of Materials and Physics, China University of Mining and Technology, Xuzhou 221116, P. R. China}

\author{K. Chang}
\email{kchang@zju.edu.cn}
\affiliation{Center for Quantum Matter, School of Physics, Zhejiang University, Hangzhou 310027, P. R. China}

\date{\today}

\begin{abstract}
The inverse Faraday effect converts the angular momentum of circularly polarized light into a quasistatic magnetization, providing a fundamental route to ultrafast optical control of magnetic fields. We develop an analytical quantum theory of this effect in intrinsic graphene, where massless Dirac fermions, electron-hole symmetry, and weak spin-orbit coupling suppress conventional population-based interpretations. We show that in intrinsic graphene, the helicity-dependent orbital magnetization arises from the off-diagonal zero-harmonic component of the second-order density matrix, rather than from intraband carrier circulation or diagonal population imbalance. We obtain closed expressions for the induced temporal surface magnetization, its finite-pulse dynamics, and the nonlocal stray magnetic field generated by the two-dimensional (2D) source. Optical helicity, frequency, pulse envelope, and beam profile independently control the response. Our results identify intrinsic graphene as an ideal platform for coherent orbital optomagnetism in 2D Dirac matter.
\end{abstract}

\maketitle

Converting optical angular momentum into electronic magnetization is a fundamental problem in nonequilibrium light--matter interaction. A paradigmatic manifestation is the inverse Faraday effect (IFE), in which circularly polarized light (CPL) generates a dc or quasistatic magnetization \cite{van1965optically,pershan1966theoretical}. Since its original formulation in nonlinear magneto-optics, the IFE has become an important route to ultrafast magnetic control because it generates an effective magnetic field without electrical contacts, coils, or magnetic proximity structures \cite{kirilyuk2010ultrafast}. This mechanism is attractive for two reasons. First, the optical helicity directly controls the sign of the response. Second, the effective field can be generated on the ultrafast picosecond or femtosecond scale to manipulate magnetization.

The microscopic origin of the IFE is nevertheless not equivalent to the classical motion of a charge in a rotating electric field. The magnetic response results from a quantum rectification process involving optical coherence, selection rules, relaxation, and the geometry of the electronic states \cite{hertel2006theory,battiato2012beyond,battiato2014quantum}. Different contributions to the nonlinear density matrix can oscillate at the optical frequency, decay as transient coherences, describe absorption, or remain at zero frequency. Only the last class can generate a genuinely quasistatic magnetization. A central theoretical problem is therefore to identify which zero-frequency density-matrix component carries the IFE and to distinguish it from harmonic and absorptive responses.

Intrinsic graphene provides an analytically controlled setting for this question. Its low-energy electrons are massless Dirac fermions described by a two-component chiral Hamiltonian \cite{castroneto2009a}. The velocity operators are proportional to Pauli matrices; consequently, the optical matrix elements are governed by the phase winding of the Dirac spinors rather than by a parabolic-band effective mass. In addition, graphene's intrinsic spin-orbit coupling is weak, so the magnetic response generated by optical driving is expected to be predominantly orbital \cite{pesin2012,hanGraphene2014}. This orbital IFE picture is consistent with quantum optomagnetic treatments of graphene \cite{abedi2023quantum}. These properties make graphene an ideal system in which the transfer of optical angular momentum to an electronic orbital degree of freedom can be treated explicitly.

The problem is closely related to the broader effort to control Dirac matter with light. Circularly polarized radiation can open Floquet gaps and generate Berry-curvature-driven Hall responses in graphene \cite{koochaki2017graphene,mciver2020light}. In graphene nanostructures and plasmonic devices, chiral electromagnetic fields can also generate strong photoinduced orbital currents and transient magnetic fields \cite{ri2019nanoscale,han2023strong,xu2024photoinduced}. These studies demonstrate that graphene can convert optical helicity into electronic orbital and topological responses. They do not, however, determine the microscopic quantum origin of the intrinsic quasistatic IFE in the clean, charge-neutral limit.

This limit matters because it differs from doped graphene. The IFE in graphene and related Dirac materials has been studied using quantum-mechanical and semiclassical approaches \cite{abedi2023quantum, tokman2020inverse}. Interband virtual transitions and interband coherence can form part of such general response descriptions. Our aim is to resolve the second-order response in the clean, charge-neutral two-band Dirac limit by both output frequency and band-index structure, and to evaluate the off-diagonal zero-harmonic contribution using the physical orbital observable defined in this work.  At charge neutrality, the valence band is filled, the conduction band is empty, and electron-hole symmetry constrains diagonal population responses \cite{castroneto2009a}. Recent quantum-kinetic theory has further shown that, in disordered 2D metals, the IFE is strongly shaped by spin-orbit coupling and impurity scattering and can even reverse sign with driving frequency, underscoring the need to establish the distinct clean-limit, interband-coherence mechanism in intrinsic graphene \cite{dzero2025inverse}. Furthermore, the diagonal orbital magnetic moment vanishes in the minimal two-band Dirac description. It is therefore unclear whether a quasistatic magnetic response survives and, if so, whether to interpret it as an interband-coherence effect.

In this Letter, we develop a quantum theory of the IFE in intrinsic monolayer graphene driven by normally incident CPL. We use a Laplace-domain density-matrix expansion to classify the second-order response according to its pole structure. We show that the IFE arises from the off-diagonal zero-harmonic component of the density matrix. In the intrinsic graphene model, this component describes coherent coupling between valence and conduction Dirac states at the same crystal momentum. We derive the helicity-dependent surface magnetization, its finite-pulse temporal envelope, and the stray magnetic field generated by a 2D magnetization. We can test the coherent response through wavelength-selective pump-probe Faraday spectroscopy \cite{zhangUltra2020}. We should differentiate between sublattice pseudospin, interband coherence, and physical orbital magnetization. Sublattice pseudospin is represented by Pauli matrices in the two-component Dirac Hamiltonian, while interband coherence is indicated by off-diagonal density matrix elements in the unperturbed Hamiltonian's eigenbasis. Neither defines physical orbital magnetization, which stems from the electronic charge-current distribution. This work examines the contribution of zero-harmonic interband coherence to the orbital response in a low-energy model, without assuming it captures all components of the inverse Faraday response in realistic graphene, which may include population, intraband, disorder, thermal, and interaction contributions.

\paragraph*{Quantum rectification and density-matrix expansion.--}
We consider normally incident CPL with a typical Gaussian spatial profile,
\begin{equation}
\boldsymbol{E}(r,t)=E_0\left(\cos\omega t\,\hat{\boldsymbol{x}}+\eta\sin\omega t\,\hat{\boldsymbol{y}}
\right){\cal R}(r)\Theta(t),\label{eq:field}
\end{equation}
with ${\cal R}(r)=e^{-r^2/r_0^2}$, the frequency $\omega$ and the strength $E_0$. $r=(x, y)$ and $\eta=\pm1$ denotes opposite helicities, $\eta= +1$ for left-handed and $\eta= -1$ for right-handed and $r_0$ is the beam waist.
The driven Hamiltonian and density matrix are written respectively as
\begin{equation}
\hat H(t)=\hat H_0+\hat V(t),\qquad \hat\rho(t)=\sum_{n=0}^{\infty}\hat\rho^{(n)}(t).
\label{eq:hamiltonian}
\end{equation}
where $\hat H_0$ is the unperturbed electronic Hamiltonian and $\hat V(t)$ is the light-matter interaction. The perturbative density matrices obey
\begin{equation}
\begin{aligned}
i\hbar\frac{\partial\hat\rho^{(0)}}{\partial t}
&=[\hat H_0,\hat\rho^{(0)}]-i\hbar\check{\Gamma}\hat\rho^{(0)}
+i\hbar\hat\rho_0\delta(t),\\i\hbar\frac{\partial\hat\rho^{(n)}}{\partial t} &=
[\hat H_0,\hat\rho^{(n)}]-i\hbar\check{\Gamma}\hat\rho^{(n)}+[\hat V(t),\hat\rho^{(n-1)}],
\end{aligned}
\label{eq:liouville}
\end{equation}
where $\check{\Gamma}$ is a phenomenological decoherence superoperator and $\hat\rho_0$ is the equilibrium density matrix.

We use the bilateral Laplace transform
\begin{equation}
\tilde f(\varepsilon)
=\int_{-\infty}^{\infty}f(t)e^{i\varepsilon t/\hbar}dt,\qquad\varepsilon=E+i\epsilon.
\label{eq:laplace}
\end{equation}
The initial condition, the convolution rule, and the transformed Liouville equation are given in the Supporting Material (SM) Sec.~S1. Following the order-notation formalism \cite{battiato2012beyond,battiato2014quantum}, we assign different order labels to operators acting on the left and right of the density matrix. We then define
\begin{equation}
\hbar^{\langle1,3\rangle}\check{\Omega} ={}^{\langle1\rangle}\hat H_0-{}^{\langle3\rangle}\hat H_0+i\hbar^{\langle1,3\rangle}\check{\Gamma}.
\label{eq:omega}
\end{equation}
The transformed hierarchy is
\begin{equation}
\hat\rho^{(n)}(\varepsilon)
=\sum_j{\cal P}\int_{-\infty}^{\infty}\frac{{}^{\langle2\rangle}
[\hat V_j,\hat\rho^{(n-1)}(\varepsilon-\varepsilon')]}{\varepsilon-\hbar^{\langle1,3\rangle}\check{\Omega}
}\frac{f_j(\varepsilon')}{2\pi\hbar}dE'.
\label{eq:general-hierarchy}
\end{equation}
The detailed order-notation derivation is given in SM Sec.~S2.

For the continuous light field, we have
\begin{equation}
\hat V(t) =\left(\hat V_+e^{i\omega t}+\hat V_-e^{-i\omega t}\right){\cal R}(r)\Theta(t),
\label{eq:V-time}
\end{equation}
with
\begin{equation}
\hat V_+=-\frac{ieE_0}{\omega}\hat v^{-\eta},~\hat V_-=\frac{ieE_0}{\omega}\hat v^{+\eta},~
\hat v^\pm=\frac{\hat v_x\pm i\hat v_y}{2}. \label{eq:V-components}
\end{equation}
The Laplace-domain interaction is
\begin{equation}
\hat V(\varepsilon)
=\frac{i\hbar\hat V_+}{\varepsilon+\hbar\omega}+\frac{i\hbar\hat V_-}{\varepsilon-\hbar\omega}.
\label{eq:V-laplace}
\end{equation}
Substituting Eq.~\eqref{eq:V-laplace} into the hierarchy, it gives
\begin{equation}
\hat\rho^{(n)}(\varepsilon)
=\sum_{j=\pm1}\frac{{}^{\langle2\rangle}[\hat V_j,\hat\rho^{(n-1)}(\varepsilon+j\hbar\omega)]
}{\varepsilon-\hbar^{\langle1,3\rangle}\check{\Omega}}.
\label{eq:rho-recursion}
\end{equation}
The first-order term is obtained as
\begin{equation}
\hat\rho^{(1)}(\varepsilon)
=\sum_{j=\pm1}\frac{i\hbar\,{}^{\langle2\rangle}[\hat V_j,\hat\rho_0]}{
(\varepsilon-\hbar^{\langle1,3\rangle}\check{\Omega})
(\varepsilon+j\hbar\omega)}.
\label{eq:rho1}
\end{equation}
The first-order poles correspond to transient and first-harmonic generation (FHG) terms. These terms either decay or oscillate and therefore do not produce a time-independent magnetization after averaging over times longer than the optical period and the dephasing time. The detailed derivation is given in SM Sec.~S3 and ~S4.

At second order, we obtain
\begin{equation}
\begin{aligned}
\hat{\rho}^{(2)}(\varepsilon)=\sum_{l,j=\pm 1}\frac{i\hbar\,^{\langle 2\rangle}\left[\hat V_l,
\frac{^{\langle 2'\rangle}[\hat V_j,\hat{\rho}_0]}{\varepsilon+l\hbar\omega-\hbar^{\langle 1',3'\rangle}\check{\Omega}}\right]}{(\varepsilon-\hbar^{\langle 1,3\rangle}\check{\Omega})
(\varepsilon+l\hbar\omega+j\hbar\omega)}.
\end{aligned}
\label{eq:rho2}
\end{equation}
Equation~\eqref{eq:rho2} contains three classes of poles,
\begin{equation}
p_a=\hbar^{\langle1,3\rangle}\check{\Omega},~
p_b=-(l+j)\hbar\omega,~p_c=\hbar^{\langle1',3'\rangle}\check{\Omega}-l\hbar\omega.
\label{eq:poles}
\end{equation}
Their inverse transforms separate the response as
\begin{equation}
\hat\rho^{(2)}
=\hat\rho_{\SHG}^{(2)}+\hat\rho_{\ABS}^{(2)}+\hat\rho_{\ZHG}^{(2)}+\hat\rho_{\TR}^{(2)}.
\label{eq:decomposition}
\end{equation}
The complete pole classification, including coincident poles and higher-order zero poles associated with absorption, is given in SM Sec.~S5 and~S6.

\begin{figure}[t]
\centering
\includegraphics[width=\columnwidth]{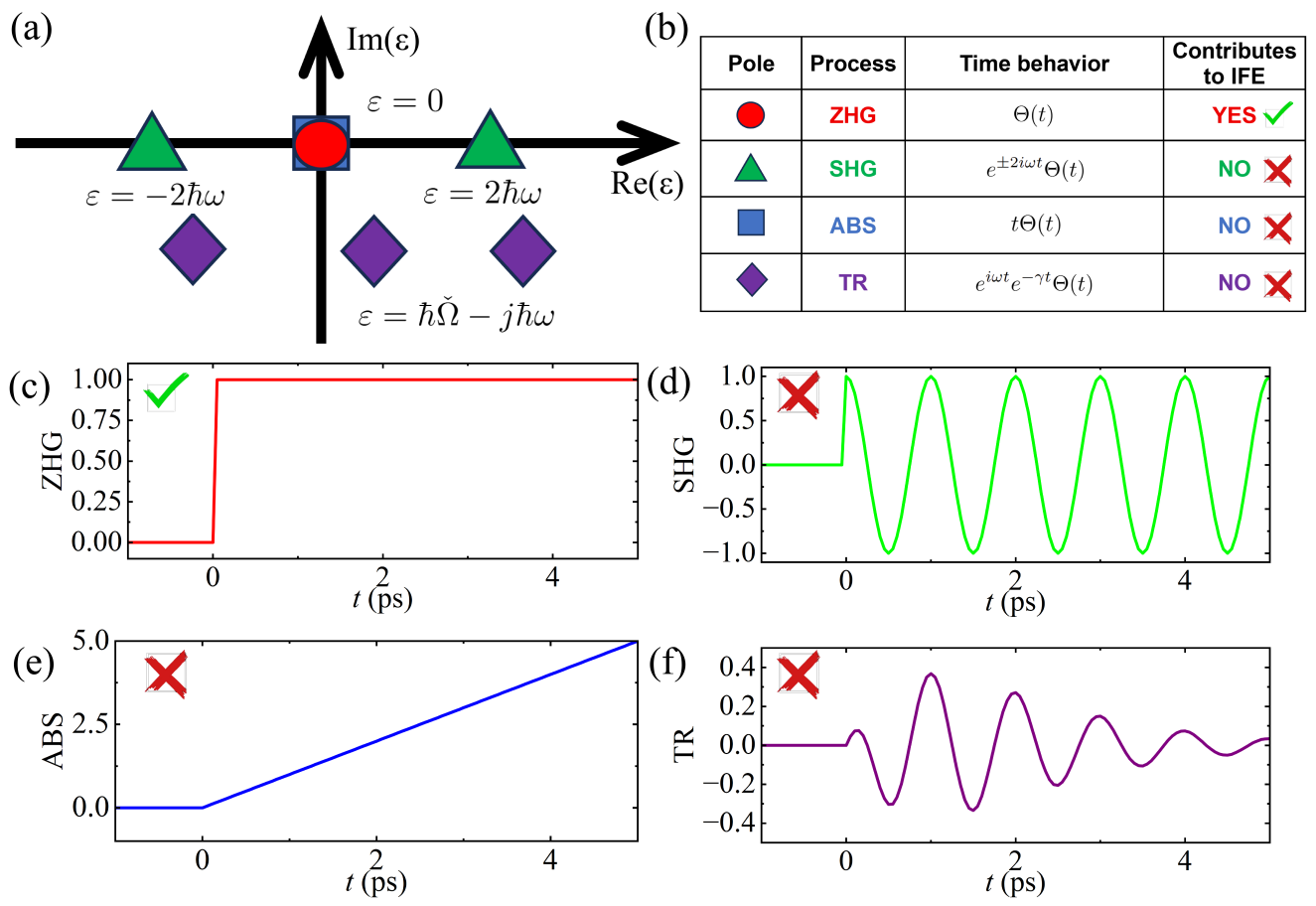}
\caption{Pole structure (a) and physical classification (b) of the second-order response. (c)--(f) The zero-harmonic (ZHG) contribution is time-independent after optical averaging. The second-harmonic (SHG) term oscillates, the absorptive (ABS) term describes secular population redistribution, and the transient (TR) terms decay with the decoherence rate. In intrinsic graphene, Fig.~\ref{fig-model} illustrates the physical picture of the diagonal zero-harmonic component.}
\label{fig-pole}
\end{figure}

The $\SHG$ terms oscillate at twice the optical frequency. The $\TR$ terms contain the decoherence poles and decay in time. The $\ABS$ terms arise from higher-order zero poles and describe secular population redistribution. By contrast, the $\ZHG$ terms contain a simple zero-frequency pole and can produce a quasistatic response. Fig.~\ref {fig-pole} summarizes the pole structure and associated temporal behaviors. In particular, the zero-harmonic pole remains after optical-cycle averaging, whereas the second harmonic and transient components do not contribute to a quasistatic IFE signal. The IFE moment is therefore
\begin{equation}
\langle\boldsymbol{M}\rangle=\mathrm{Tr}\left[\hat{\boldsymbol{M}}\cdot \hat\rho_{\ZHG}^{(2)}
\right]. \label{eq:moment-trace}
\end{equation}
It is useful to separate this quantity into
\begin{align}
\boldsymbol{M}_{\rm IFEo}&=\sum_{n\neq m}\mel{m}{\hat{\boldsymbol{M}}}{n}\mel{n}{\hat\rho_{\ZHGo}^{(2)}}{m}, \nonumber\\\boldsymbol{M}_{\rm IFEda}
&=\sum_n\mel{n}{\hat{\boldsymbol{M}}}{n}\mel{n}{\hat\rho_{\ZHGda}^{(2)}}{n},\nonumber\\
\boldsymbol{M}_{\rm IFEdb}&=\sum_n\mel{n}{\hat{\boldsymbol{M}}}{n}\mel{n}{\hat\rho_{\ZHGdb}^{(2)}}{n}.
\label{eq:moment-components}
\end{align}
The matrix-element expressions of all three components are given in Sec.~S6 and S7 of SM for intermediate states $m,n$. We separate the second-order density matrix into diagonal population changes and off-diagonal interband coherences in the eigenbasis of $\hat H_0$. We evaluate the off-diagonal zero-harmonic component's contribution to the physical orbital response by contracting that density matrix with the corresponding projected orbital-moment matrix elements. Thus, ``interband coherence'' refers to the density-matrix band structure, whereas orbital magnetization refers to the physical observable defined above. The statement that a particular contribution vanishes or survives must be made at the level of this physical observable, not from the pseudospin texture alone.

\paragraph*{Dirac-band origin of the IFE.--}
We now specialize to intrinsic monolayer graphene. Near the Dirac point, we have
\begin{equation}
\hat H_0=v_F\boldsymbol{\sigma}\cdot\hat{\boldsymbol p},\qquad \mathcal{E}_\pm=\pm\hbar v_F|\boldsymbol{k}|,
\label{eq:dirac}
\end{equation}
where $\mathcal{E}_\pm$ are  eigenstates for conduction $(+)$ and valence $(-)$ bands, respectively, with $v_F\simeq10^6$~m/s, the Pauli matrix $\boldsymbol{\sigma}$, and momentum $\hat{\boldsymbol p}=\hbar {\boldsymbol k}$. Fig.~\ref{fig-model} illustrates the physical picture. A normally
incident Gaussian CPL beam writes a localized out-of-plane surface magnetization on graphene [Fig.~\ref{fig-model}(a)]. Microscopically, the optical field coherently couples valence and conduction Dirac states at the same crystal momentum and generates the off-diagonal zero-harmonic component $\hat\rho_{\ZHGo}^{(2)}$ [Fig.~\ref{fig-model}(b)]. The corresponding eigenstates are
\begin{equation}
\ket{\boldsymbol{k},+}=\frac{1}{\sqrt2}\begin{pmatrix}e^{-i\theta_k/2}\\e^{i\theta_k/2}\end{pmatrix},
\qquad\ket{\boldsymbol{k},-}=\frac{1}{\sqrt2}\begin{pmatrix}e^{-i\theta_k/2}\\-e^{i\theta_k/2}
\end{pmatrix},
\label{eq:spinors}
\end{equation}
where $\theta_k=\arctan(k_y/k_x)$ \cite{dong2008}. The half-angle phase in Eq.~\eqref{eq:spinors} represents the winding of the Dirac pseudospin. This winding, rather than a conventional orbital trajectory, determines the circular optical matrix elements.

The velocity operators follow from the Heisenberg equation with $\hat v_x=v_F\sigma_x$ and $\hat v_y=v_F\sigma_y$, therefore
\begin{equation}
\hat v^+=v_F
\begin{pmatrix}
0&1\\0&0
\end{pmatrix},
\qquad\hat v^-=v_F
\begin{pmatrix}
0&0\\1&0
\end{pmatrix}.
\label{eq:velocity-matrices}
\end{equation}
The complete matrix elements of $\hat v^\pm$ and $\hat V_\pm$ in the conduction-valence basis are listed in SM Sec.~S8. Their angular factors are fixed by $\theta_k$ and encode the helicity-dependent selection rules.

Because intrinsic graphene has weak spin-orbit coupling \cite{pesin2012,hanGraphene2014}, we focus on the orbital part of the magnetic response. We stress, however, that the sublattice pseudospin of the Dirac Hamiltonian is not the physical electron spin and that its angular texture is not, by itself, a definition of the orbital magnetic moment.

For a finite two-dimensional sample of area $\mathcal{A}$, we define the $z$ component of the orbital magnetic moment from the physical charge sheet-current density as
\begin{equation}
\hat{\mathcal M}^{\rm orb}_z =\frac{1}{2}\int_{\mathcal{A}} d^2r\,
\left[ x\,\hat j_y(\boldsymbol r) - y\,\hat j_x(\boldsymbol r) \right].
\label{eq:physical-orbital-moment-main}
\end{equation}
Here $\boldsymbol r=(x,y)$ is the in-plane position, $\hat j_x(\boldsymbol r)$ and $\hat j_y(\boldsymbol r)$ are the
in-plane components of the physical charge-sheet current-density operator, and the current is obtained from the microscopic electromagnetic coupling using the charge-current convention of the Hamiltonian. For a normalized density operator $\hat\rho$, the area-averaged $z$-component of the orbital magnetization is
\begin{equation}
M_z^{\rm orb} = \frac{1}{\mathcal{A}} \operatorname{Tr}\!\left[
\hat{\mathcal M}^{\rm orb}_z\hat\rho \right],
\label{eq:physical-orbital-magnetization}
\end{equation}
where the trace is over the sample's electronic states. The finite-sample definition above is not, by itself, a prescription for evaluating bulk orbital magnetization from extended Bloch states. For a periodic system, a bulk calculation requires an appropriate periodic-system formulation or a specified limiting procedure from finite samples. In this work, we discuss the low-energy projection and its relation to the angular dependence of the Dirac spinors in Sec.~S9 of the SM. Sublattice pseudospin and interband coherence are not, by themselves, physical orbital magnetic moments; their contribution to the response is determined by the projection of the current-defined observable.

For clarity, in the valley convention used, the low-energy Hamiltonian Eq.~\eqref{eq:dirac} with the two inequivalent valleys $\tau$, Pauli matrices act on the two sublattice components. The in-plane orbital angular-momentum operator is $\hat L_z=\hat x\hat p_y-\hat y\hat p_x$, with $\hat{\boldsymbol p}=-i\hbar\nabla$ the momentum operator. Within this continuous Dirac model, the generator of simultaneous rotations of real space and sublattice pseudospin is
\begin{equation}
\hat J_{z,\tau} =\hat L_z+\tau\frac{\hbar}{2}\sigma_z, \qquad [\hat J_{z,\tau},\hat H_{0,\tau}]=0.
\label{eq:dirac-conserved-angular-main}
\end{equation}
Here $\hat J_{z,\tau}$ is the $z$ component of the total rotation generator in the Dirac model; it does not imply that sublattice pseudospin is physical spin or orbital magnetization.

The following interband relation refers to the projection and wave-packet representation specified in Sec.~S9 of the SM, and is not an independent definition of the physical orbital moment:
\begin{equation}
\left\langle +,\boldsymbol k\middle|\hat L_z\middle|-,\boldsymbol k\right\rangle_{\rm proj}
= -\tau\frac{\hbar}{2} \left\langle +,\boldsymbol k\middle|\sigma_z\middle|-,\boldsymbol k\right\rangle.
\label{eq:dirac-L-matrix-main}
\end{equation}
Here $|+,\boldsymbol k\rangle$ and $|-,\boldsymbol k\rangle$ denote, respectively, the conduction- and valence-band Dirac spinors at the same momentum. The phase of an interband matrix element depends on the relative phase convention of the two band spinors. In the spinor gauge used there, for valley $\tau=+1$,
$\langle +,\boldsymbol k|\sigma_z|-,\boldsymbol k\rangle=1$; therefore the projected relation gives
$\langle +,\boldsymbol k|\hat L_z|-,\boldsymbol k\rangle_{\rm proj} =-\hbar/2$. This gauge-specific projected angular-momentum matrix element is not, on its own, the matrix element of the current-defined physical orbital-moment operator $\hat{\mathcal M}^{\rm orb}_z$.

Moreover, using the covariant angular derivative in the Dirac-band basis and the
spinor gauge of Eq.~\eqref{eq:spinors}, the projected interband orbital-moment matrix
elements are
\begin{equation}
\begin{aligned}
&\mel{-}{M_z}{+}=\mel{+}{M_z}{-}=-\frac{\mu_B}{2},\\&
\mel{+}{M_z}{+}=\mel{-}{M_z}{-}=0.
\label{eq:Mz_elements}
\end{aligned}
\end{equation}
These results have a direct physical consequence. Within the projected orbital-moment model adopted here, the diagonal orbital-moment matrix elements vanish. Consequently, the calculated quasistatic contribution is carried by the off-diagonal zero-harmonic interband coherence. The magnetic response must be carried by an off-diagonal interband coherence as
{\begin{equation}
\boldsymbol{M}_{\rm IFE}=\sum_{n\neq m}\mel{m}{\hat{\boldsymbol{M}}}{n}\mel{n}{\hat\rho_{\ZHGo}^{(2)}}{m}.
\label{eq:coherent-IFE}
\end{equation}
The projected orbital-moment operator in Eq.~\eqref{eq:physical-orbital-magnetization} and the trace in Eq.~\eqref{eq:coherent-IFE} follow the standard gauge-invariant Bloch orbital-magnetization formulation \cite{xiao2010berry, tazuke2025}; individual interband matrix elements are gauge-dependent. The symmetry argument, Raman-type quantum pathways, and the equivalent Dirac-pseudospin picture of this off-diagonal zero-harmonic response are given in SM Sec.~S9. 

\begin{figure}[t]
\centering
\includegraphics[width=0.94\columnwidth]{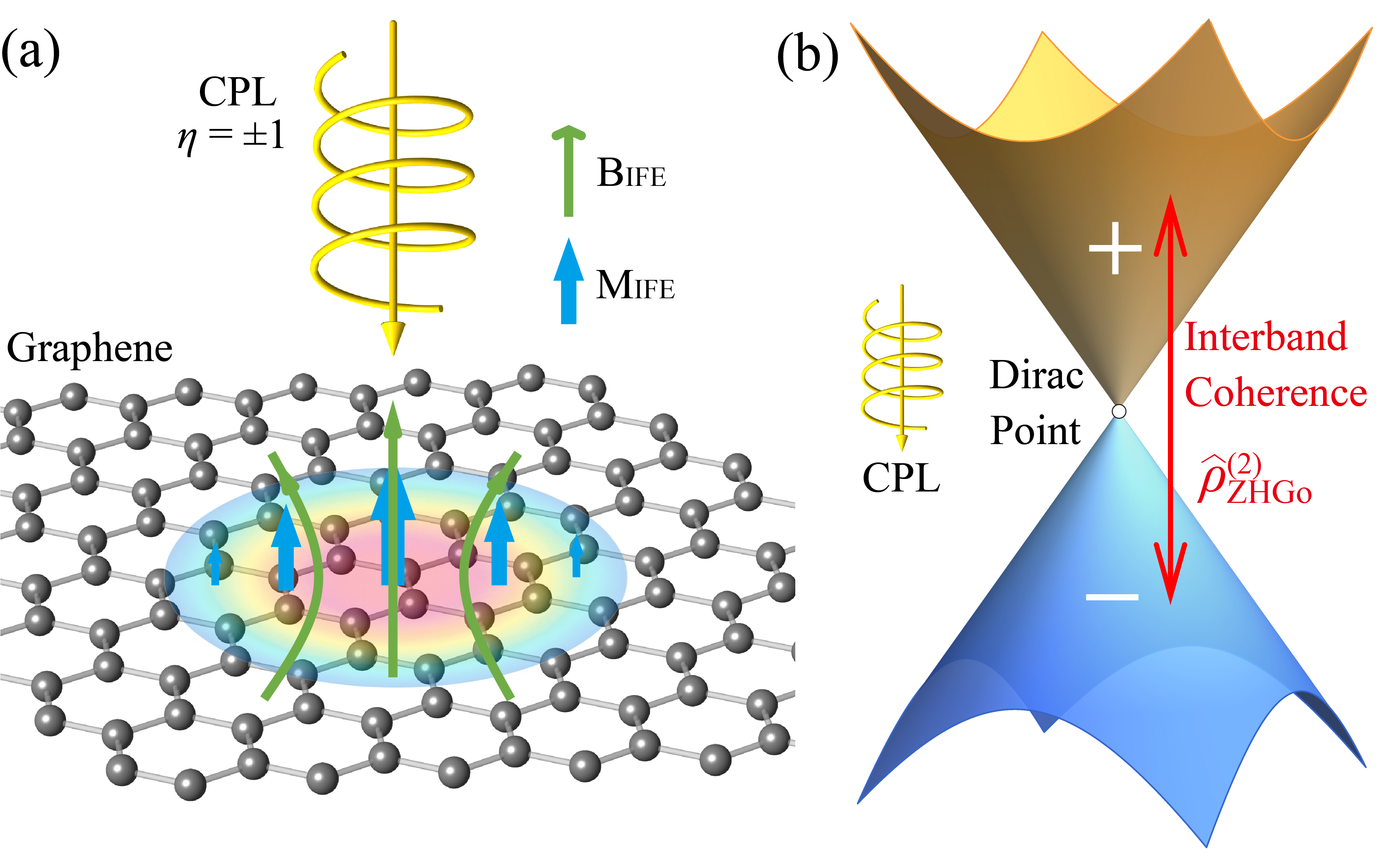}
\caption{Microscopic model of the IFE in intrinsic graphene. (a) A Gaussian CPL spot induces an out-of-plane surface magnetization on graphene. The corresponding $\boldsymbol B_{\rm IFE}$ is a stray field produced by this localized 2D source. (b) Circularly polarized light couples valence and conduction Dirac states at fixed momentum and generates the off-diagonal zero-harmonic coherence $\hat\rho_{\ZHGo}^{(2)}$.}
\label{fig-model}
\end{figure}

At zero temperature and zero chemical potential, the valence band is filled, and the conduction band is empty. Vertical optical transitions preserve crystal momentum, so the intermediate state in the second-order response is restricted to the two-band subspace at fixed $\boldsymbol{k}$. Including spin $g_s=2$ and valley degeneracy $g_v=2$, the coherent moment strength becomes
\begin{equation}
\begin{aligned}
&M^z_{\rm IFE}=\frac{g_sg_v\mu_B}{8\pi^2}{\cal R}^2(r)\int_0^\infty dk
\int_0^{2\pi}d\theta_k\frac{k}{2\hbar v_Fk+i\hbar\Gamma}
\\&\times\left[\frac{\eta e^2E_0^2v_F^2/(2\omega^2)}{-2\hbar v_Fk-\hbar\omega-i\hbar\Gamma}
\right. \left.-\frac{\eta e^2E_0^2v_F^2/(2\omega^2)}{-2\hbar v_Fk+\hbar\omega-i\hbar\Gamma}\right].
\end{aligned}
\label{eq:moment-integral}
\end{equation}
The two terms represent the two helicity-resolved virtual optical paths. Their difference is odd under $\eta\rightarrow-\eta$ and is the microscopic origin of the helicity-dependent response. We consider charge-neutral monolayer graphene at $T=0$ near both Dirac valleys, with wave vectors measured from each valley center and restricted to $k \cdot a\ll1$, where $a$ is the lattice constant. We neglect spin--orbit coupling, so spin provides a passive factor $g_s=2$. We include both valleys $\tau$ explicitly. We take optical transitions as vertical and neglect intervalley scattering as a model assumption. Phenomenological interband dephasing does not constitute a microscopic treatment of disorder or transport. Moreover, we discuss gauge covariance of the projected orbital response in S9 of the SM.

Using the weak-dephasing limit and the Sokhotski--Plemelj prescription, the integral gives
\begin{equation}
\boldsymbol{M}_{\rm IFE}(r)=-\eta M_0{\cal R}^2(r)\hat{\boldsymbol z},\qquad
M_0=\frac{e^2E_0^2\mu_B}{8\hbar^2\omega^2}.\label{eq:main-result}
\end{equation}
The result is local in the optical intensity but nonlocal in its observable magnetic field, as discussed below. It immediately implies
\begin{equation}
\boldsymbol{M}_{\rm IFE}(\eta)=-\boldsymbol{M}_{\rm IFE}(-\eta).
\label{eq:helicity-odd}
\end{equation}
For left-handed light, the magnetization points along $-\hat{\boldsymbol z}$, while reversing the helicity reverses the magnetization direction. 

At fixed electric-field amplitudes, Eq.~\eqref{eq:main-result} gives $M_0\propto\omega^{-2}$. This dependence arises from the two velocity-gauge vertices, each contributing a factor of $\omega^{-1}$. It does not imply an unlimited low-frequency enhancement. At sufficiently low frequency, disorder, finite temperature, residual carriers, and intraband dynamics become important, so the clean interband theory must then be supplemented by transport terms. Within its regime of validity, however, Eq.~\eqref{eq:main-result} identifies the THz range as a favorable window for the intrinsic coherent IFE.

At fixed field amplitude, the formal $M_0\propto E_0^2/\omega^2$ scaling applies only in the weak-drive, clean interband regime; it cannot be extrapolated to $\omega\to0$. Comparing the field-induced momentum shift $\delta k=eE_0/(\hbar\omega)$ with the resonant interband momentum $k_\omega=\omega/(2v_F)$ gives the perturbative condition \cite{Ornigotti2023, Semnani2019}
\begin{equation}
\alpha_{\rm D}\equiv \frac{2e v_F E_0}{\hbar\omega^2}\ll1.
\label{eq:main-validity-alpha-D}
\end{equation}
The clean, nearly neutral interband regime further requires
\begin{equation}
\max(\hbar\Gamma_{+-},k_BT,|\mu|) \ll\hbar\omega\ll E_\Lambda,
\label{eq:main-validity-interband}
\end{equation}
where $\Gamma_{+-}$ is the interband dephasing rate and $E_\Lambda$ is the energy cutoff of the linear Dirac model. At $T=0$ in the clean limit, interband transitions are Pauli-blocked for $\hbar\omega<2|\mu|$, where $\mu$ is the chemical potential. The low-frequency crossover is not a universal sharp cutoff: its form depends on dephasing, temperature, doping, pulse duration, and intraband transport \cite{Semnani2019}. Outside this regime, the present result describes the clean interband contribution, not the complete IFE response. Please see S9 of the SM for details.

Using representative parameters from graphene experiments \cite{han2023strong}, $f=\omega/2\pi=3.5~\mathrm{THz}, \quad\gamma=1~\mathrm{THz}, \quad I_0=500~\mathrm{W/\mu m^2},
\quad r_0=500~\mathrm{nm}$, we estimate a surface magnetization scale of order $M^z_{\rm IFE}\sim 10^{3}\mu_B/{\mu \rm m}^2$. This is comparable in magnitude to magnetic moments reported for defect-dominated graphene nanocrystal laminates \cite{sepioni2010limits}. The comparison is only a scale estimate: static defect magnetism and the present helicity-controlled interband-coherence response have different physical origins.

\paragraph*{Ultrafast laser pulses.--}
To investigate temporal control, we use a circularly polarized pulse with finite rise and fall times \cite{battiato2014quantum},
\begin{align}
\hat V(r,t)={}&\sum_{j=\pm1}e^{ij\omega t}\hat V_j{\cal R}(r)\left[
(1-e^{-\gamma t})\Theta(t)\right.\nonumber\\
&\left.-(1-e^{-\gamma(t-T)})\Theta(t-T)\right],
\label{eq:pulse-potential}
\end{align}
where $T$ is the pulse duration and $\gamma$ controls the temporal edge. We derive the Laplace transform of Eq.~\eqref{eq:pulse-potential}, the pole decomposition, and the inverse transform in SM Sec.~S10.

The off-diagonal ZHG component has a factorized form,
\begin{equation}
\hat\rho_{\ZHGo}^{(2)}(t)=\sum_{j=\pm1}\frac{{}^{\langle2\rangle}\left[\hat V_j,\dfrac{{}^{\langle2'\rangle}
[\hat V_{-j},\hat\rho_0]}{\hbar^{\langle1',3'\rangle}\check{\Omega}
-j\hbar\omega}\right]}{\hbar^{\langle1,3\rangle}\check{\Omega}}{\cal T}(t),
\label{eq:pulse-density}
\end{equation}
with the time envelope
\begin{align}
&{\cal T}(t)=(1-e^{-2\gamma t})\Theta(t)-\Theta(t-T)\nonumber\\
&\times \left[1+e^{-\gamma t}-e^{-\gamma(t-T)}-e^{-2\gamma(t-T/2)}\right].
\label{eq:envelope}
\end{align}
Consequently, we obtain
\begin{equation}
\boldsymbol{M}_{\rm IFE}^{\rm pulse}(r,t)=-\eta M_0{\cal R}^2(r){\cal T}(t)\hat{\boldsymbol z}.
\label{eq:pulse-magnetization}
\end{equation}
Equation~\eqref{eq:pulse-magnetization} makes the roles of the optical control parameters explicit, as illustrated in Fig.~\ref{fig-waveform}. The left panels display the carrier-frequency optical perturbation, whereas the right panels show the rectified zero-harmonic magnetization envelope.
Increasing the pulse duration $T$ extends the time over which the induced magnetization persists [Figs.~\ref{fig-waveform}(a) and \ref{fig-waveform}(b)]. The switching rate $\gamma$ controls the rise and decay times [Figs.~\ref{fig-waveform}(c) and \ref{fig-waveform}(d)]. Increasing the frequency suppresses the magnetization amplitude, consistent with $M_0\propto\omega^{-2}$ [Figs.~\ref{fig-waveform}(e) and \ref{fig-waveform}(f)]. The absence of carrier-frequency oscillations in the magnetic response confirms that the IFE is a rectified zero-harmonic coherence rather than a cycle-resolved carrier orbit.

\begin{figure}[t]
\centering
\includegraphics[width=\columnwidth]{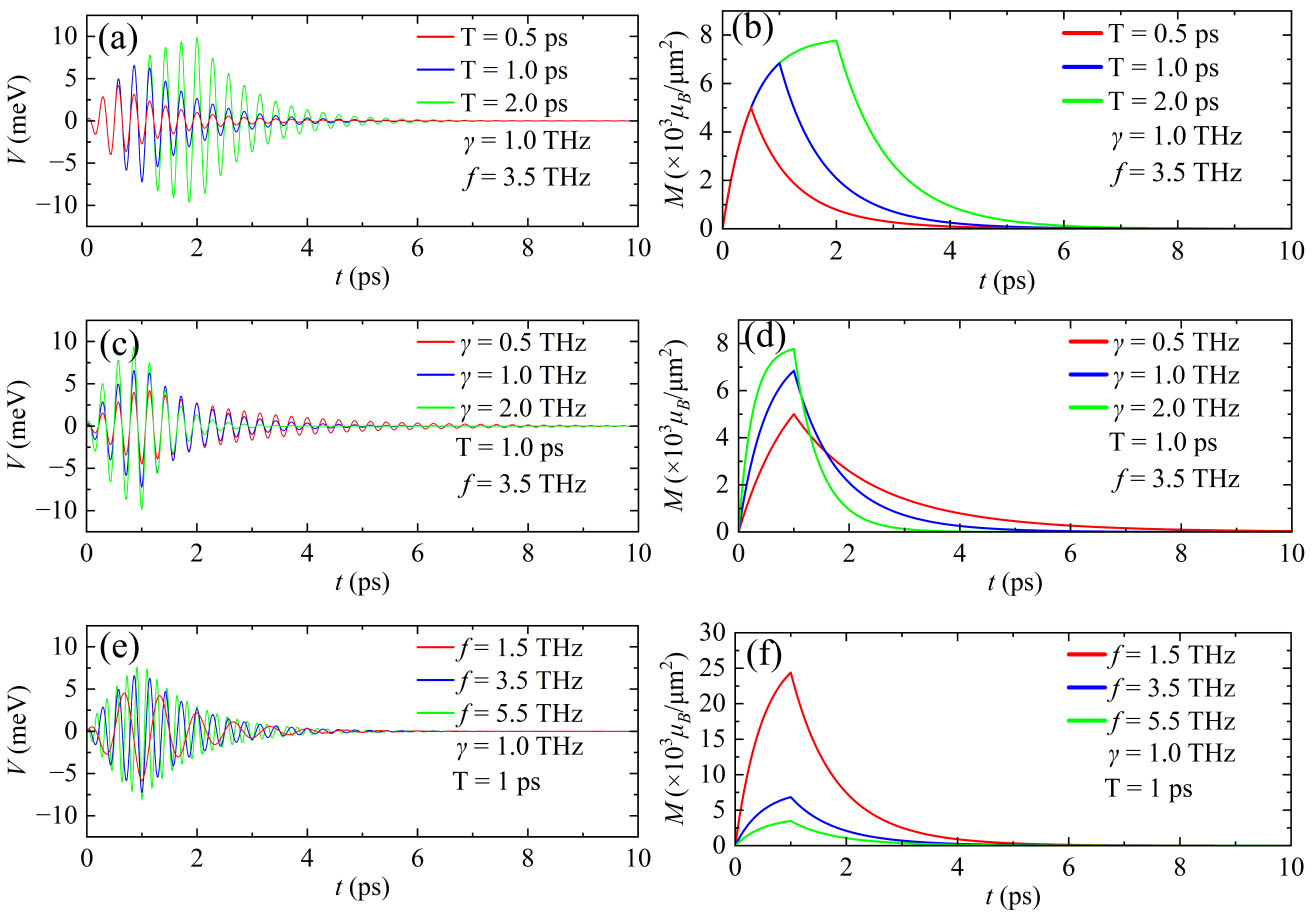}
\caption{Temporal response under pulsed CPL. The optical perturbations (a), (c) and (e) contain carrier-frequency oscillations, whereas the corresponding IFE (b), (d) and (f) follow the zero-harmonic envelope ${\cal T}(t)$. The pulse duration $T$ controls the persistence of the magnetization, $\gamma$ controls the switching rate, and the frequency dependence follows $M_0\propto\omega^{-2}$ at fixed field amplitude.}
\label{fig-waveform}
\end{figure}

For a longer pulse duration $T$, the response reaches the steady value in Eq.~\eqref{eq:main-result} over an extended time interval. A sharper pulse edge, corresponding to larger $\gamma$, produces a faster switching response. These features differ from the absorptive contribution, which is associated with a secular change in diagonal occupations. Here the temporal response is determined by the coherent zero-harmonic component and remains helicity odd throughout the pulse.

\paragraph*{2D magnetization and stray field.--}
The induced magnetization is confined to the graphene plane and should be interpreted as a surface magnetization density. For a Gaussian beam, one has 
\begin{equation}
m_z(r)=-\eta M_0e^{-2r^2/r_0^2}.
\label{eq:surface-moment}
\end{equation}
Its zeroth-order Hankel transform is
\begin{equation}
\tilde m(k)=\int_0^\infty m_z(r)J_0(kr)r\,dr=-\eta M_0\frac{r_0^2}{4}e^{-k^2r_0^2/8}.
\label{eq:hankel}
\end{equation}
This expression shows that the beam waist determines the magnetic source's spatial spectrum. A tightly focused spot contains larger in-plane wave vectors and produces a more localized field, whereas a broader spot produces a longer-ranged spatial distribution.

The magnetic field $\boldsymbol{B}$ outside the graphene plane is obtained from the scalar magnetic potential, $\boldsymbol H=-\nabla\Phi_m$, satisfying Laplace's equation away from the sheet. The potential discontinuity at the graphene plane is set by the surface magnetization:
\begin{equation}
\nabla^2\Phi_m=0, \qquad \Phi_m(r,0^+)-\Phi_m(r,0^-)=m_z(r).
\label{eq:potential}
\end{equation}
For an axisymmetric source,
\begin{equation}
\Phi_m(r,z)=\frac{\mathrm{sgn}(z)}{2}\int_0^\infty\tilde m(k)e^{-k|z|}J_0(kr)k\,dk.
\label{eq:potential-integral}
\end{equation}
with $\mathcal{J}_n(x)=\frac{1}{\pi}\int_0^{\pi}\cos(n\theta-x\sin \theta)d\theta$ being the first kind of Bessel function. The resulting induction is
\begin{align}
B_z(r,z)&=\frac{\mu_0}{2}\int_0^\infty\tilde m(k)e^{-k|z|}J_0(kr)k^2\,dk,
\nonumber\\B_r(r,z)&=\frac{\mathrm{sgn}(z)\mu_0}{2}\int_0^\infty\tilde m(k)e^{-k|z|}J_1(kr)k^2\,dk,
\label{eq:magnetic-field}
\end{align}
with $\mu_0$ being the vacuum permeability. The field is therefore not confined to the graphene sheet. It extends into the surrounding space and has the nonlocal character expected for the stray field of a localized out-of-plane surface moment. For a pulsed excitation,
\begin{equation}
\boldsymbol B(r,z,t)
=\boldsymbol B(r,z){\cal T}(t).\label{eq:pulsed-field}
\end{equation}
The resulting three-dimensional (3D) field distribution is shown in Fig.~\ref{fig-magnetic-field}. The field is largest near the center of the illuminated region and near the graphene plane
[Figs.~\ref{fig-magnetic-field}(a) and \ref{fig-magnetic-field}(c)]. The vertical cross section and field streamlines reveal the dipolar, nonlocal character of the stray field generated by the finite surface magnetization [Figs.~\ref{fig-magnetic-field}(b) and \ref{fig-magnetic-field}(d)]. Reversing the helicity reverses the surface magnetization and hence the complete magnetic-field pattern. The magnetic field strength at the center reaches $B_0 = 0.29$ mT at $f = 100$ GHz.

The ultrafast magneto-optical measurement would provide a prospective test of the predicted helicity-dependent response. A helicity-odd signal alone would not uniquely identify interband coherence; frequency-, time-, and doping-dependent measurements would be needed to distinguish competing contributions. The local optical intensity determines $m_z(r)$, but the field results from convolution with the magnetostatic kernel. The strongest field is therefore concentrated near the optical axis and the graphene plane, while field lines extend above and below the sheet. Reversing $\eta$ reverses $m_z(r)$ and consequently reverses every component of $\boldsymbol B$.

\begin{figure}[t]
\centering
\includegraphics[width=\columnwidth]{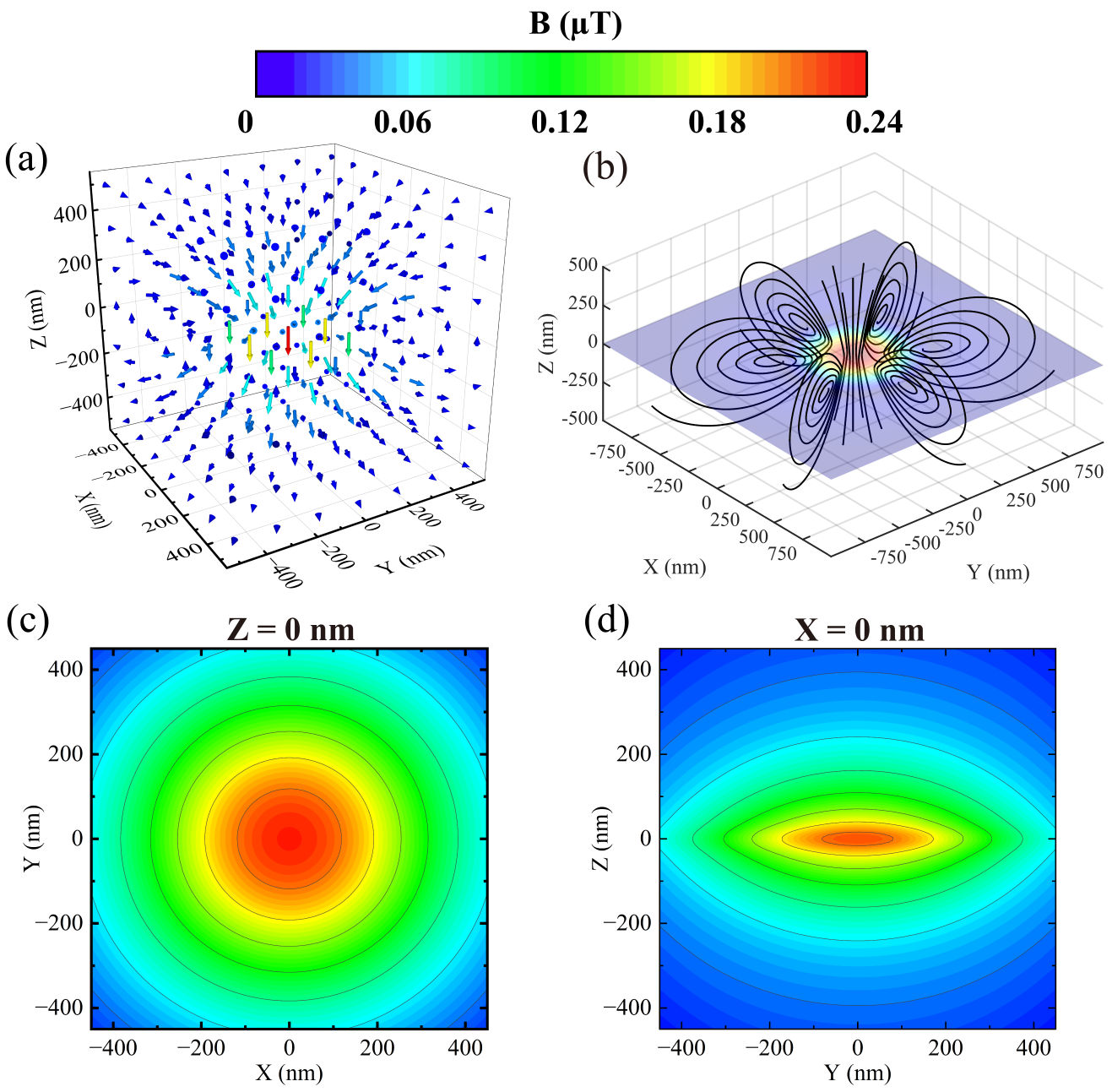}
\caption{Stray magnetic field generated by the optically induced Gaussian surface magnetization. (a) 3D vector distribution diagram of the magnetic field. (b) 3D streamline diagram of the magnetic field; the illustration shows the magnetic field magnitude in the $Z=0$ plane. (c) and (d) the map of the magnetic field in the $Z=0$ plane and $X=0$ plane, respectively. The color represents the magnetic field magnitude.}
\label{fig-magnetic-field}
\end{figure}

The separability between time and space is a useful consequence of the perturbative clean-limit theory. The pulse envelope controls when the magnetization is established and removed, while $r_0$ controls where the magnetic field is concentrated. It shows that helicity controls the sign, and frequency controls the amplitude. This suggests independent optimization of temporal switching and spatial localization in a graphene optomagnetic device.

The results presented here are limited to the clean, charge-neutral, weak-field Dirac model and to the specific current-based orbital observable mentioned earlier. Within this model, the second-order response includes a helicity-dependent, off-diagonal zero-harmonic component. This component differs distinctly in its frequency and band-index structure from both the second-harmonic response and from diagonal population redistribution. However, this separation does not imply that interband coherence is absent from more general interband optical excitation theories, nor does it mean that other channels are absent in realistic graphene samples \cite{tazuke2025}. In particular, at low frequencies or with finite doping levels, a transport treatment is necessary in addition to the interband density-matrix response discussed here.  The value of this calculation lies in establishing an analytically controlled reference limit, where the origin of the ultrafast coherence response is clear and unambiguous. The present theory isolates the intrinsic coherent interband contribution. It excludes finite chemical potential, residual thermal carriers, edge currents, disorder-assisted absorption, plasmonic enhancement, many-body renormalization, and strong-field Floquet reconstruction. These effects may become important in realistic samples and can add intraband, dissipative, or interaction-induced contributions.

In conclusion, we have analyzed the helicity-dependent inverse-Faraday response of intrinsic graphene within a clean, charge-neutral, weak-field Dirac model. In this limit, the quasistatic second-order density matrix contains an off-diagonal zero-harmonic interband-coherence contribution. Its contribution to the orbital response is obtained by contracting the coherence with
the corresponding projected, current-defined orbital-moment matrix element; the coherence alone is not a magnetic observable. Within the assumptions and observable specified in this work, this channel gives the surface-magnetization profile
\begin{equation}
\boldsymbol{M}_{\rm IFE}(r) = -\eta\frac{e^2E_0^2\mu_B}{8\hbar^2\omega^2} {\cal R}^2(r)\,\hat{\boldsymbol z}.
\label{eq:conclusion-result}
\end{equation}
The pulse envelope and beam profile determine the temporal response and spatial stray-field profile, respectively. The result provides an analytically controlled reference for this coherent contribution
to the IFE, establishes intrinsic graphene as a minimal platform for coherent orbital optomagnetism and provides an analytically controlled benchmark for the IFE in graphene and related 2D Dirac materials. The absolute magnetization scale is conditional on the current-based projection used to define the orbital observable. 

\begin{acknowledgments}
 This work is supported by Jiangsu Key Laboratory of Low-Dimensional Material Physics and Physical State Manipulation Open Projects Fund (No. LDMP2602), the Science and Technology Program of Xuzhou (KC25001) and by the National Natural Science Foundation of China (Grant Nos. 12374079, 11604380).
\end{acknowledgments}

\bibliography{ref}

@article{van1965optically,
  title={Optically-induced magnetization resulting from the inverse {F}araday effect},
  author={Van der Ziel, JP and Pershan, Peter S and Malmstrom, LD},
  journal={Physical Review Letters},
  volume={15},
  number={5},
  pages={190},
  year={1965},
  publisher={APS},
  doi = {10.1103/PhysRevLett.15.190}
}

@article{pershan1966theoretical,
  title={Theoretical discussion of the inverse {F}araday effect, Raman scattering, and related phenomena},
  author={Pershan, PS and Van der Ziel, JP and Malmstrom, LD},
  journal={Physical Review},
  volume={143},
  number={2},
  pages={574},
  year={1966},
  publisher={APS},
  doi = {10.1103/PhysRev.143.574}
}

@article{kirilyuk2010ultrafast,
  title={Ultrafast optical manipulation of magnetic order},
  author={Kirilyuk, Andrei and Kimel, Alexey V and Rasing, Theo},
  journal={Reviews of Modern Physics},
  volume={82},
  number={3},
  pages={2731--2784},
  year={2010},
  publisher={APS},
  doi = {10.1103/RevModPhys.82.2731}
}

@article{hertel2006theory,
  title={Theory of the inverse {F}araday effect in metals},
  author={Hertel, Riccardo},
  journal={Journal of Magnetism and Magnetic Materials},
  volume={303},
  number={1},
  pages={L1--L4},
  year={2006},
  publisher={Elsevier},
  doi = {10.1016/j.jmmm.2005.10.225}
}

@article{battiato2012beyond,
  title={Beyond linear response theory for intensive light-matter interactions: Order formalism and ultrafast transient dynamics},
  author={Battiato, Marco and Barbalinardo, G and Carva, Karel and Oppeneer, Peter M},
  journal={Physical Review B},
  volume={85},
  number={4},
  pages={045117},
  year={2012},
  publisher={APS},
  doi = {10.1103/PhysRevB.85.045117}
}

@article{battiato2014quantum,
  title={Quantum theory of the inverse {F}araday effect},
  author={Battiato, Marco and Barbalinardo, G and Oppeneer, Peter M},
  journal={Physical Review B},
  volume={89},
  number={1},
  pages={014413},
  year={2014},
  publisher={APS},
  doi = {10.1103/PhysRevB.89.014413}
}

@article{pesin2012,
  title = {Spintronics and {P}seudospintronics in {G}raphene and {T}opological {I}nsulators},
  author = {Pesin, Dmytro and MacDonald, Allan H.},
  year = 2012,
  month = may,
  journal = {Nature Materials},
  volume = {11},
  number = {5},
  pages = {409--416},
  issn = {1476-1122, 1476-4660},
  doi = {10.1038/nmat3305},
  urldate = {2026-06-10},
  langid = {english}
}

@article{hanGraphene2014,
  title = {Graphene {S}pintronics},
  author = {Han, Wei and Kawakami, Roland K. and Gmitra, Martin and Fabian, Jaroslav},
  year = 2014,
  month = oct,
  journal = {Nature Nanotechnology},
  volume = {9},
  number = {10},
  pages = {794--807},
  issn = {1748-3387, 1748-3395},
  doi = {10.1038/nnano.2014.214},
  urldate = {2026-06-10},
  langid = {english}
}

@article{koochaki2017graphene,
  title={Graphene superlattices in strong circularly polarized fields: {C}hirality, {B}erry phase, and attosecond dynamics},
  author={Koochaki Kelardeh, Hamed and Apalkov, Vadym and Stockman, Mark I},
  journal={Physical Review B},
  volume={96},
  number={7},
  pages={075409},
  year={2017},
  publisher={APS},
  doi = {10.1103/PhysRevB.96.075409}
}

@article{mciver2020light,
  title={Light-induced anomalous {H}all effect in graphene},
  author={McIver, James W and Schulte, Benedikt and Stein, F-U and Matsuyama, Toru and Jotzu, Gregor and Meier, Guido and Cavalleri, Andrea},
  journal={Nature physics},
  volume={16},
  number={1},
  pages={38--41},
  year={2020},
  publisher={Nature Publishing Group UK London},
  doi = {10.1038/s41567-019-0698-y}
}

@article{ri2019nanoscale,
  title={Nanoscale magnetization and third-order nonlinearity by the plasmon-induced inverse {F}araday effect in graphene-covered semiconductors},
  author={Ri, Chol-Song and Im, Song-Jin and Pae, Ji-Song and Ho, Kum-Song and Han, Yong-Ha and Herrmann, Joachim},
  journal={Physical Review B},
  volume={100},
  number={15},
  pages={155404},
  year={2019},
  publisher={APS},
  doi = { 10.1103/PhysRevB.100.155404}
}

@article{han2023strong,
  title={Strong transient magnetic fields induced by {THz}-driven plasmons in graphene disks},
  author={Han, Jeong Woo and Sai, Pavlo and But, Dmytro B and Uykur, Ece and Winnerl, Stephan and Kumar, Gagan and Chin, Matthew L and Myers-Ward, Rachael L and Dejarld, Matthew T and Daniels, Kevin M and others},
  journal={Nature Communications},
  volume={14},
  number={1},
  pages={7493},
  year={2023},
  publisher={Nature Publishing Group UK London},
  doi = {10.1038/s41467-023-43412-x}
}

@article{xu2024photoinduced,
  title={Photoinduced currents and inverse {F}araday effect in graphene quantum dots},
  author={Xu, Zi-Yang and Xie, Hang},
  journal={Physical Review B},
  volume={110},
  number={8},
  pages={085425},
  year={2024},
  publisher={APS},
  doi = {10.1103/PhysRevB.110.085425}
}

@article{tokman2020inverse,
  title={Inverse {F}araday effect in graphene and {W}eyl semimetals},
  author={Tokman, ID and Chen, Qianfan and Shereshevsky, IA and Pozdnyakova, VI and Oladyshkin, Ivan and Tokman, Mikhail and Belyanin, Alexey},
  journal={Physical Review B},
  volume={101},
  number={17},
  pages={174429},
  year={2020},
  publisher={APS},
  doi = {10.1103/PhysRevB.101.174429}
}

@article{dong2008,
  title = {Quantum and {T}ransport {C}onductivities in Monolayer Graphene},
  author = {Dong, H. M. and Xu, W. and Zeng, Z. and Lu, T. C. and Peeters, F. M.},
  year = 2008,
  month = jun,
  journal = {Physical Review B},
  volume = {77},
  number = {23},
  pages = {235402},
  issn = {1098-0121, 1550-235X},
  doi = {10.1103/PhysRevB.77.235402},
  urldate = {2024-10-31},
  langid = {english}
}

@article{sepioni2010limits,
  title={Limits on intrinsic magnetism in graphene},
  author={Sepioni, M and Nair, RR and Rablen, S and Narayanan, J and Tuna, F and Winpenny, R and Geim, AK and Grigorieva, IV},
  journal={Physical Review Letters},
  volume={105},
  number={20},
  pages={207205},
  year={2010},
  publisher={APS},
  doi = {10.1103/PhysRevLett.105.207205}
}

@article{castroneto2009a,
  title = {The {E}lectronic {P}roperties of {G}raphene},
  author = {Castro Neto, A. H. and Guinea, F. and Peres, N. M. R. and Novoselov, K. S. and Geim, A. K.},
  year = 2009,
  month = jan,
  journal = {Reviews of Modern Physics},
  volume = {81},
  number = {1},
  pages = {109--162},
  issn = {0034-6861, 1539-0756},
  doi = {10.1103/RevModPhys.81.109},
  urldate = {2026-07-24},
  langid = {english}
}

@article{dzero2025inverse,
  author  = {Dzero, Maxim},
  title   = {Inverse {F}araday {E}ffect in Disordered Two-Dimensional Electronic Systems},
  journal = {Physical Review Letters},
  volume  = {135},
  pages   = {116902},
  year    = {2025},
  doi     = {10.1103/PhysRevLett.135.116902}
}

@article{zhangUltra2020,
  title = {Ultrafast {Terahertz} {Magnetometry}},
  author = {Zhang, Wentao and Maldonado, Pablo and Jin, Zuanming and Seifert, Tom S. and Arabski, Jacek and Schmerber, Guy and Beaurepaire, Eric and Bonn, Mischa and Kampfrath, Tobias and Oppeneer, Peter M. and Turchinovich, Dmitry},
  year = 2020,
  month = aug,
  journal = {Nature Communications},
  volume = {11},
  number = {1},
  pages = {4247},
  issn = {2041-1723},
  doi = {10.1038/s41467-020-17935-6},
  urldate = {2026-08-22},
  langid = {english}
}

@article{abedi2023quantum,
doi = {10.1088/1361-6455/ace395},
url = {https://doi.org/10.1088/1361-6455/ace395},
year = {2023},
month = {jul},
publisher = {IOP Publishing},
volume = {56},
number = {14},
pages = {145502},
author = {Abedi, Sina and Hamed Majedi, A},
title = {Quantum optomagnetics in graphene},
journal = {Journal of Physics B: Atomic, Molecular and Optical Physics}
}

@article{xiao2010berry,
  author  = {Xiao, Di and Chang, Ming-Che and Niu, Qian},
  title   = {Berry phase effects on electronic properties},
  journal = {Rev. Mod. Phys.},
  volume  = {82},
  pages   = {1959},
  year    = {2010},
  doi     = {10.1103/RevModPhys.82.1959}
}

@article{tazuke2025,
  title = {Formulation of the Orbital Magnetic Moment in Multiorbital Tight-Binding Models: {{Application}} to the Inverse {{Faraday}} Effect},
  author = {Tazuke, Kosuke and Morimoto, Takahiro and Kitamura, Sota},
  year = 2025,
  month = oct,
  journal = {Physical Review B},
  volume = {112},
  number = {15},
  pages = {155134},
  issn = {2469-9950, 2469-9969},
  doi = {10.1103/gmnv-cwvr},
  urldate = {2026-01-30},
  langid = {english}
}

@article{Ornigotti2023,
  author  = {Ornigotti, Marco and Carvalho, David N. and Biancalana, Fabio},
  title   = {Nonlinear optics in graphene: theoretical background and recent advances},
  journal = {La Rivista del Nuovo Cimento},
  volume  = {46},
  number  = {6},
  pages   = {295--380},
  year    = {2023},
  doi     = {10.1007/s40766-023-00043-8}
}

@article{Semnani2019,
  author  = {Semnani, Behrooz and Jago, Roland and Safavi-Naeini, Safieddin
             and Majedi, Hamed and Malic, Ermin and Tassin, Philippe},
  title   = {Anomalous optical saturation of low-energy {D}irac states in graphene
             and its implication for nonlinear optics},
  journal = {2D Materials},
  volume  = {6},
  number  = {3},
  pages   = {031003},
  year    = {2019},
  doi     = {10.1088/2053-1583/ab1dea}
}

\end{document}